\documentclass[preprint,useAMS,usenatbib]{mn2e}
\usepackage{amsmath}
\usepackage{natbib}
\usepackage{graphicx}
\usepackage{amssymb}
\usepackage{amsfonts}
\usepackage{verbatim}
\usepackage{times}
\usepackage[usenames]{color}

\newcommand{\apjl}{ApJL}
 
\newcommand{\apj}{ApJ} 
\newcommand{\aj}{AJ}

\newcommand{\mnras}{MNRAS}
\newcommand{\nat}{Nature}

\newcommand{\araa}{ARA\&A}
\newcommand{\aap}{A\&A}

\def\gtsim {\lower .1ex\hbox{\rlap{\raise .6ex\hbox{\hskip .3ex
        {\ifmmode{\scriptscriptstyle >}\else
                {$\scriptscriptstyle >$}\fi}}}
        \kern -.4ex{\ifmmode{\scriptscriptstyle \sim}\else
                {$\scriptscriptstyle\sim$}\fi}}}

\newcommand{\eg}{{\rm e.g.}}
\newcommand{\ie}{{\rm i.e.}}

\newcommand{\dr}{\ensuremath{\Delta r}}
\newcommand{\dzm}{\ensuremath{\Delta z_\mathrm{max}}}
\newcommand{\zm}{\ensuremath{z_\mathrm{max}}}
\newcommand{\rform}{\ensuremath{R_\mathrm{form}}}

\newcommand{\msun}{\ensuremath{M_{\odot}}}
\newcommand{\sfe}{\ensuremath{c^{\star}}}
\newcommand{\zd}[1]{\ensuremath{z_\mathrm{d}={#1}\ \mathrm{pc}}}
\newcommand{\mdr}{\ensuremath{{\Delta r_\mathrm{med}}}}
\newcommand{\drt}{\ensuremath{{\Delta r_\mathrm{80}}}}

\newcommand{\rinit}{\ensuremath{r_\mathrm{i}}}
\newcommand{\rfin}{\ensuremath{r_\mathrm{f}}}
\newcommand{\dcirc}{\ensuremath{\Delta\epsilon}}
\newcommand{\cir}{\ensuremath{\epsilon}}
\newcommand{\rg}{\ensuremath{R_\mathrm{g}}}

\newcommand{\be}{\begin{equation}}
\newcommand{\ee}{\end{equation}}
\newcommand{\beqa}{\begin{eqnarray}}
\newcommand{\eeqa}{\end{eqnarray}}

\def\Mo{{\rm M_\odot}}

\def\kpc{\ \ensuremath{\rm kpc}}
\def\pc{\ \ensuremath{\rm pc}}

\def\kms{\ \ensuremath{ {\rm km}\,{\rm s}^{-1}}}
\def\kmssq{\ \ensuremath{ {\rm km}^2\,{\rm s}^{-2}}}
\def\LCDM{$\Lambda$CDM}

\begin{document}

\title[Radial Mixing in Galactic Disks]{Radial Mixing in Galactic Disks: The
Effects of Disk Structure and Satellite Bombardment}
\author[Bird, Kazantzidis, \& Weinberg]{Jonathan C.
Bird\thanks{E-mail:bird@astronomy.ohio-state.edu}, Stelios Kazantzidis, \& David
H. Weinberg\\
Department of Astronomy and the Center for Cosmology and
  Astro-Particle Physics, The Ohio State University, Columbus,~OH~43210}

\pagerange{\pageref{firstpage}--\pageref{lastpage}} \pubyear{2011}

\date{\today}

\maketitle

\label{firstpage}

\begin{abstract}
We use a suite of numerical simulations to investigate the mechanisms
and effects of radial migration of stars in disk galaxies like the
Milky Way (MW).  An isolated, collisionless stellar disk with a
MW-like scale-height shows only the radial ``blurring'' expected from
epicyclic orbits. Reducing the disk thickness or adding gas to the
disk substantially increases the level of radial migration, induced by
interaction with transient spiral arms and/or a central bar. We also
examine collisionless disks subjected to gravitational perturbations
from a cosmologically motivated satellite accretion history. In the
perturbed disk that best reproduces the observed properties of the MW,
20\% of stars that end up in the solar annulus $7\kpc < R < 9\kpc$
started at $R < 6\kpc$, and 7\% started at $R > 10\kpc$. This level of
migration would add considerable dispersion to the age-metallicity
relation of solar neighborhood stars. In the isolated disk models, the
probability of migration traces the disk's radial mass profile, but in
perturbed disks migration occurs preferentially at large radii, where
the disk is more weakly bound. The orbital dynamics of migrating
particles are also different in isolated and perturbed disks:
satellite perturbations drive particles to lower angular momentum for
a given change in radius. Thus, satellite perturbations appear to be a
distinct mechanism for inducing radial migration, which can operate in
concert with migration induced by bars and spiral structure.  We
investigate correlations between changes in radius and changes in
orbital circularity or vertical energy, identifying signatures that
might be used to test models and distinguish radial migration
mechanisms in chemo-dynamical surveys of the MW disk.
\end{abstract}


\section{Introduction}\label{sec:intro}

Under the influence of gravity, disk galaxies are expected to assemble
in an ``inside-out'' fashion: stars form first from high-density gas
in the central region of the galaxy where the potential is deepest,
and subsequently at increasing galacto-centric radii
\citep[\eg][]{Larson76}. An immediate consequence of this formation
scenario is that stars born at the same time and in the same region of
a galaxy should have similar chemical compositions. However,
observations in our Galaxy suggest that these initial conditions are
not maintained. \citet{Wielen96} argued that the Sun was substantially
more metal rich than nearby solar age stars and the local interstellar
medium (ISM). A recent recalibration of the Geneva Copenhagen Survey
(GCS) using the infrared flux method finds no discrepancy with solar
age stars \citep{Casagrande11}, but even modern studies confirm that
the age-metallicity relationships (AMRs) of field and solar
neighborhood stars are characterized by higher dispersions than
expected \citep{Edvardsson93,Nordstrom04, Holmberg07,
Casagrande11}. In addition, simple chemical evolution models that
divide the Galaxy into independently evolving concentric annuli
predict many more low metallicity G-dwarfs in our region of the disk
compared to those observed, a discrepancy known as ``the local G-dwarf
problem'' \citep[][]{vandenbergh62, Schmidt63}. Evidence seemingly in
contradiction to standard galaxy chemical evolution theory is not
limited to our own Galaxy. Metallicity gradients in disk galaxies are
shallower than predicted by classical models
\citep[\eg,][]{Magrini07}. \citet{Ferguson01} and \citet{Ferguson07}
find unexpectedly old stellar populations on nearly circular orbits in
the outskirts of M31 and M33, respectively. The outermost regions of
NGC300 and NGC7739 show flattened or positive abundance gradients with
radius \citep{Vlajic09,Vlajic11}. These perplexing observations cannot
be readily explained within the confines of classic galaxy formation
models.

A natural explanation for the observational challenges above arises if
the present day radii of many stars could be significantly different
from their birth radii. One difficulty in establishing radial
migration as a common phenomenon, from a dynamical standpoint, lies in
finding a mechanism that can cause a substantial fraction of stars to
migrate several kiloparsecs while retaining the observed approximately
circular orbits. In a seminal paper, \citet[][hereafter
SB02]{Sellwood02} investigated the relationship between changes in
stellar angular momentum and disk heating. They found that radial
migration is a ubiquitous process in spiral galaxies; stars naturally
migrate (change angular momentum) as they resonantly interact with
transient spiral waves. Stars in corotational resonance (CR) with said
waves are scattered without heating the disk and maintain their nearly
circular orbits (unlike Lindblad resonance (LR) scattering). In the
present paper, we examine radial migration in simulations of disk
galaxies. Our experiments include galactic disks evolved both in
isolation and under the action of infalling satellites of the type
expected in the currently favored cold dark matter (CDM) paradigm of
hierarchical structure formation
\citep[\eg,][]{Peebles82,Blumenthal_etal84}.  The latter set of
experiments were presented in the studies of \citet[][hereafter
K08]{Kazantzidis08} and \citet{Kazantzidis09} and were utilized to
investigate the {\it generic} dynamical and morphological signatures
of galactic disks subject to bombardment by CDM substructure.

Inspired by SB02, several groups have recently investigated the
potential role of radial mixing in the chemical and dynamical
evolution of disk galaxies. \citet{Schonrich09} presented the first
chemical evolution model to incorporate radial migration.  The rate at
which stars migrate via the SB02 mechanism is left as a free parameter
constrained by the metallicity distribution function (MDF) of solar
neighborhood stars in the GCS \citep{Nordstrom04}. Their model
successfully reproduced, within systematic uncertainties, the observed
age-metallicity distribution of stars in the GCS \citep{Holmberg07}
and the observed correlation between tangential velocity and abundance
pattern described by \citet{Haywood08}. However, there is partial
degeneracy between the magnitude of radial migration and other
parameters in the model such as star-formation rates and gas inflow
characteristics. Furthermore, it is unclear whether the level of
migration required to fit the data is consistent with theoretical
expectations .

Numerical simulations have confirmed the occurrence of radial
migration under a variety of conditions. \citet{Roskar08a, Roskar08b}
studied the migration of stars in a simulation of an isolated Milky
Way (MW)-sized stellar disk formed from the cooling of a
pressure-supported gas cloud in a $10^{12} \msun$ dark matter halo.
In their simulations, some older stars radially migrated to the
outskirts of the disk while maintaining nearly circular orbits,
forming a population akin to that observed in M31 and M33
\citep{Ferguson01, Ferguson07}. \citet{Roskar08b} found that $\sim
50\%$ of all stars in the solar neighborhood were not born \emph{in
situ}; this is a natural explanation for the observed dispersion in
the AMR and solar neighborhood metallicity distribution function
(MDF).

More recently, \citet{Quillen09} investigated radial migration in a
stellar disk perturbed by a low-mass ($\sim 5\times10^9\ \msun$)
orbiting satellite. Their numerical simulations integrated test
particle orbits in a static galactic potential and highlighted the
fact that mergers and perturbations from satellite galaxies and
subhalos can induce stellar radial mixing. Although informative, test
particle simulations in a static isothermal potential will not capture
all the relevant physics of the process of stellar radial migration in
disk galaxies; the interactions between the gravitational
perturbations and the self-gravity of the disk are essential to a
detailed analysis of the phenomenon. In our paper, we expand on the
analysis of \citet{Quillen09} by investigating radial migration using
fully self-consistent numerical simulations both with and without
satellite bombardment.

There are now several established phenomena that can cause a star to
populate a region of the disk different from its birth radii. Stars on
elliptical orbits maintain their guiding center and angular momentum
(modulo asymmetries in the potential) but can be found over the range
in galacto-centric radius defined by their pericenter and
apocenter. Changing a star's angular momentum, and hence its guiding
center, requires direct scattering or a resonant interaction with
transient patterns in the disk. The local encounters of stars with
molecular clouds \citep[\eg][]{Spitzer53} or Lindblad resonance (LR)
scattering between stars and spiral waves (\eg\ SB02) both change
stellar guiding centers (albeit to a relatively small degree) and
increase the random motions of stars over time, ``blurring'' the
disk. Stars scattered at CR with spiral waves can change their guiding
centers by several kiloparsecs without increasing the amplitude of
their radial motion. For any single spiral wave, SB02 predict that
stars are scattered on each side of the CR, ``churning'' the contents
of the disk\footnote{Blurring and churning are the terms proposed by
\citet{Schonrich09} to describe these distinct aspects of radial
migration.}.  Stars may undergo several encounters with transient
spiral waves throughout their lifetimes. While SB02 investigate all
resonant interactions between spiral waves and stars, we will refer to
this special case of CR as the ``SB02 mechanism''\footnote{In the
recent radial migration studies of secularly evolved simulations,
significant migration is almost always attributed to CR
scattering.}. As SB02 note, migration due to spiral waves can be
described by blurring and churning regardless of how the waves arise
(satellites, \eg, could induce spiral structure that would lead to
migration described by SB02). Simulations have shown that other
transient wave patterns internal to a galaxy, such as those resulting
from bar propagation, can produce resonance overlap with existing
spiral patterns and induce radial migration \citep{Minchev10,
Minchev11}. Orbiting satellites, external to the galaxy and discussed
above, will have a complex interaction with the disk as they provide a
means of direct scattering over a large area and also induce spiral
modes in the disk. In this work, we aim to characterize radial
migration induced by satellite bombardment and compare its effects on
the stellar disk with those observed in secularly evolved galaxies.

Our investigation complements earlier and ongoing radial migration
studies. We perform a simulation campaign, including numerical
experiments of isolated disk galaxies with different scale heights and
gas fractions, which in turn lead to different levels of spiral
structure. For the first time, we examine the effect of satellite
bombardment on radial migration utilizing simulations where galactic
disks are subjected to a cosmologically motivated satellite accretion
history. Via a comparative approach, we determine how the magnitude
and efficiency of radial migration depend on input physics, establish
correlations between orbital parameters and migration, and present
evidence that each of the three migration mechanisms is distinct in
the examined parameter space. These characteristics lead to possible
observational signatures that may constrain the relative importance of
each migration mechanism in the Milky Way.
%----------------------------
\section{Methods}
\label{sec:methods}
%----------------------------
\subsection{Isolated Disk Models}
\label{sec:isolated_disks}

We employ the method of \citet{Widrow_Dubinski05} to construct
numerical realizations of self-consistent, multicomponent disk
galaxies. These galaxy models consist of an exponential stellar disk,
a Hernquist bulge \citep{Hernquist90}, and a \citet[][hereafter
NFW]{Navarro_etal96} dark matter halo. They are characterized by $15$
free parameters that may be tuned to fit a wide range of observational
data for actual galaxies including the MW and M31. The
\citet{Widrow_Dubinski05} models are derived from three-integral,
composite distribution functions and thus represent self-consistent
equilibrium solutions to the coupled Poisson and collisionless
Boltzmann equations. Owing to their self-consistency, these galaxy
models are ideally suited for investigating the complex dynamics
involved in the process of radial mixing. The
\citet{Widrow_Dubinski05} method has been recently used in a variety
of numerical studies associated with instabilities in disk galaxies,
including the dynamics of warps and bars
\citep{Dubinski_Chakrabarty09,Dubinski_etal09}, the gravitational
interaction between galactic disks and infalling satellites
(\citealt{Gauthier_etal06}; K08; \citealt{Purcell_etal09};
\citealt{Kazantzidis09}), and the transformation of disky dwarfs to
dwarf spheroidal galaxies under the action of tidal forces from a
massive host \citep{Kazantzidis11}. We refer the reader to
\citet{Widrow_Dubinski05} for an overview of all relevant parameters
and a detailed description of this technique.

For the majority of numerical experiments in the present study, we
employ model ``MWb'' in \citet{Widrow_Dubinski05}, which satisfies a
broad range of observational constraints on the MW galaxy.
Specifically, the stellar disk has a mass of $M_{\rm disk} = 3.53
\times 10^{10}\ \Mo$, a radial scale length of $R_d=2.82\kpc$, and a
sech$^2$ scale height of $z_d=400\pc$. We note that the adopted value
for the scale height is consistent with that inferred for the old,
thin stellar disk of the MW
\citep[\eg,][]{Kent_etal91,Juric_etal08}. The equivalent exponential
scale height is approximately $200$ pc, but the sech$^2$ vertical
distribution is more accurate.  The bulge has a mass and a scale
radius of $M_b=1.18 \times 10^{10}\ \Mo$ and $a_b=0.88\kpc$,
respectively. The NFW dark matter halo has a tidal radius of
$R_h=244.5 \kpc$ to keep the total mass finite (K08), a mass of
$M_h=7.35 \times 10^{11}\ \Mo$, and a scale radius of
$r_h=8.82\kpc$. The total circular velocity of the galaxy model at the
solar radius, $R_{\odot} \simeq 8 \kpc$, is $V_c(R_\odot)=234.1\kms$,
and the Toomre disk stability parameter is equal to $Q = 2.2$ at
$R=2.5 R_d$. We note that direct numerical simulations of the
evolution of model MWb in isolation confirm its stability against bar
formation for $10$~Gyr.

We wish to address the dependence of radial mixing in isolated disk
galaxies upon initial disk thickness and the presence of gas in the
galactic disk. Because of their smaller ``effective'' Toomre $Q$
stability parameter \citep[\eg,][]{Romeo11}, thinner disks
are more prone to gravitational instabilities and thus yield stronger
and better defined spiral structure. Therefore, we might expect them
to cause enhanced radial mixing compared to their thicker
counterparts.  Similarly, because radiative cooling in the gas damps
its random velocities, the presence of gas is associated with
long-lived spiral structure in disks
\citep[\eg,][]{Carlberg_Freedman85}, which may act to increase the
amount of radial mixing. Correspondingly, we initialize several
additional disk galaxy models.

The first modified galaxy model was constructed with the same
parameter set as MWb but with a scale height $2$ times smaller
($z_d=200\pc$). Except for disk thickness and vertical velocity
dispersion, all of the other gross properties of the three galactic
components of this model are within a few percent of the corresponding
ones for MWb. Not surprisingly, being more prone to gravitational
instabilities, this model develops a bar inside
$\sim 5\kpc$ at time $t \sim 1.2$~Gyr. The second set of modified
galaxy models are the same as MWb except for the fact that a fraction
$f_g$ of the mass of the initial stellar disk is replaced by gas.
Thus, the resulting gaseous component is constructed with the same
initial density distribution as the stellar disk. In the present
study, we employ two values for the gas fraction, $f_g=0.2$ and
$f_g=0.4$, and the adopted methodology will be described in detail in
a forthcoming paper (Kazantzidis et al.  2011, in preparation).
Briefly, the construction of the gas disk is done by assuming that its
vertical structure is governed by hydrostatic equilibrium and by
computing the potential and the resulting force field for the radially
varying density structure of the gas component.  By specifying the
polytropic index and the mean molecular weight of the gas and using a
tree structure for the potential calculation, the gas azimuthal
streaming velocity is determined by the balance between gravity and
centrifugal and pressure support.

\subsection{Perturbed Disk Models}
\label{sec:perturbed_disks}

In the standard CDM paradigm of hierarchical structure formation,
galaxies grow via continuous accretion of smaller satellite
systems. To investigate the effect of accretion events on radial
mixing in galactic disks, we have also analyzed $N$-body simulations
of the gravitational interaction between a population of dark matter
subhalos and disk galaxies. These simulations were first presented by
K08, and we summarize them briefly here.

K08 performed high-resolution, collisionless $N$-body simulations to
study the response of the galactic model MWb discussed above to a
typical {\LCDM}-motivated satellite accretion history. The specific
merger history was derived from cosmological simulations of the
formation of Galaxy-sized CDM halos and involved six dark matter
satellites with masses, orbital pericenters, and tidal radii of $7.4
\times 10^{9} \lesssim M_{\rm sat}/M_{\odot} \lesssim 2 \times
10^{10}$, $r_{\rm peri} \lesssim 20\kpc$, and $r_{\rm tid} \gtrsim 20
\kpc$, respectively, crossing the disk in the past $\sim 8$~Gyr.  As a
comparison, the Large Magellanic Cloud has an estimated {\it total}
present mass of $\sim 2 \times 10^{10}\ M_{\odot}$
\citep[\eg,][]{Schommer_etal92,Mastropietro_etal05}, so the masses of
the perturbing dark matter satellites correspond to the upper limit of
the mass function of observed satellites in the Local Group. K08
modeled satellite impacts as a sequence of encounters. Specifically,
starting with the first subhalo, they included subsequent systems at
the epoch when they were recorded in the cosmological simulation. We
note that although K08 followed the accretion histories of host halos
since $z \sim 1$, when time intervals between subhalo passages were
larger than the timescale needed for the disk to relax after the
previous interaction, the next satellite was introduced immediately
after the disk had settled from the previous encounter. Each satellite
was removed from the simulation once it reached its maximum distance
from the disk after crossing.  This approach was dictated by the
desire to minimize computational time and resulted in a total
simulation time of $\sim 2.5$~Gyr instead of $\sim 8$~Gyr that would
formally correspond to $z=1$.

K08 and \citet{Kazantzidis09} found that these accretion events
severely perturbed the galactic disk of model MWb without destroying
it and produced a wealth of distinctive morphological and dynamical
signatures on its structure and kinematics. In this paper, we will
investigate the magnitude of radial mixing induced in disk model MWb
by these accretion events. As in the case of isolated disk models, it
is worthwhile to examine the dependence of radial mixing upon disk
thickness. For this purpose, we employed the thinner galaxy model with
a scale height of $z_d=200\pc$ described above and repeated the
satellite-disk encounter simulations of K08.

\begin{figure*}
  \includegraphics[width=6in]{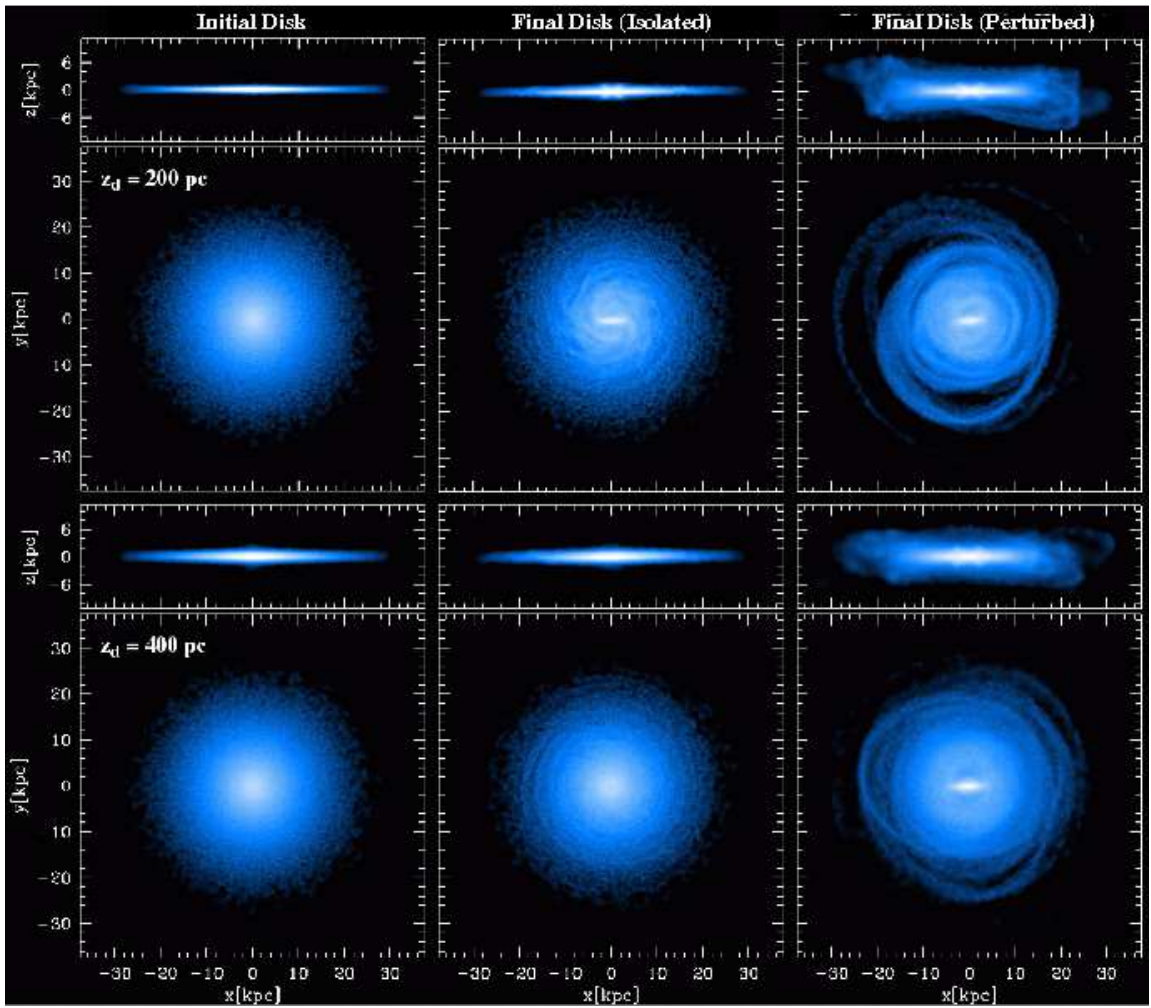}
  \caption{Surface density maps of the stellar distributions of
    galactic disks with different initial scale-heights $z_d$. Maps
    are presented only for the collisionless experiments and include
    the initial ({\it left panels}), final isolated ({\it middle
    panels}), and final perturbed disks ({\it right panels}). Each
    panel includes the face-on ({\it bottom panels}) and edge-on ({\it
    upper panels}) distributions of disk stars.  Particles are
    color-coded on a logarithmic scale, with hues ranging from blue to
    white indicating increasing stellar density.  Local density is
    calculated using an SPH smoothing kernel of $32$ neighbors.  The
    density ranges from $7 \times 10^5$ to $2 \times 10^9 \Mo/\kpc^2$.
    \label{fig:sims}}
\end{figure*}

\subsection{Numerical Parameters}
\label{sec:num_param}

All collisionless numerical simulations discussed in this paper were
carried out with the multi-stepping, parallel, tree $N$-body code
PKDGRAV \citep{Stadel01}. The hydrodynamical simulations were
performed with the parallel TreeSPH $N$-body code GASOLINE
\citep{Wadsley_etal04}. In the gasdynamical experiments, we include
atomic cooling for a primordial mixture of hydrogen and helium, star
formation and (thermal) feedback from supernovae. Our star formation
recipe follows that of \citet{Stinson_etal06}, which is based on that
of \citet{Katz92}. Gas particles in cold and dense regions which are
simultaneously parts of converging flows spawn star particles with a
given efficiency $c^{\star}$ at a rate proportional to the local
dynamical time. Feedback from supernovae is treated using the
blast-wave model described in \citet{Stinson_etal06}, which is based
on the analytic treatment of blastwaves described in
\citet{McKee_Ostriker77}. In our particular applications, gas
particles are eligible to form stars if their density exceeds 0.1
atoms/cm$^3$ and their temperature drops below $T_{\rm
max}=1.5\times10^4$~K, and the energy deposited by each Type-II
supernova into the surrounding gas is $4\times10^{50}$~erg.  We note
that this choice of parameters and numerical techniques is shown to
produce realistic disk galaxies in cosmological simulations
\citep{Governato_etal07}. Lastly, in an attempt to investigate the
effect of star formation efficiency on radial mixing, for each of the
gas fractions $f_g$ above, we adopted two different values for
$c^{\star}$, namely $c^{\star}=0.05$ (which reproduces the slope and
normalization of the observed Schmidt law in isolated disk galaxies)
and a lower value of $c^{\star}=0.01$.

All $N$-body realizations of the disk galaxy models contain
$N_d=10^{6}$ particles of $m_d=3.53\times10^4\msun$ in the disk;
$N_b=5\times10^{5}$, $m_b=2.36\times10^4\msun$ in the bulge; and
$N_h=2\times10^{6}$, $m_d=3.675\times10^5\msun$ in the dark matter
halo. The gravitational softening lengths for the three components
were set to $\epsilon_d=50 \pc$, $\epsilon_b=50 \pc$, and
$\epsilon_h=100 \pc$, respectively. These are ``equivalent Plummer''
softenings; the force softening is a cubic spline. In the
hydrodynamical simulations of isolated galaxies, gaseous disks were
represented with $N_g=2\times10^{5}$ and $N_g=4\times10^{5}$ for gas
fractions of $f_g=0.2$ and $f_g=0.4$, respectively. In these cases,
the gravitational softening length for the gas particles was set to
$\epsilon_g = 50 \pc$. All numerical simulations of isolated and
perturbed disks were analyzed at $2.5$~Gyr. This time-scale
corresponds to approximately $17.5$ orbital times at the disk
half-mass radius. For evaluating the impact of accretion events, it is
fairest to compare isolated and perturbed disks after the same amount
of integration, but we may underestimate the total amount of mixing in
the isolated (and, to a lesser extent the perturbed) models. In the
future, we plan to evolve selected simulations for the full $8$ Gyr
interval since $z=1$, but such simulations will require more than
three times the computational resources used here.

Exchange of angular and linear momentum between the infalling
satellites and the disk tilt the disk plane and cause the disk center
of mass to drift from its initial position at the origin of the
coordinate frame (K08; \citealt{Kazantzidis09}). Therefore, we
calculate \dr\ in the satellite-disk encounter experiments after
removing the global displacement and tilt of the disk galaxy by
determining the principal axes of the total disk inertia tensor and
rotating our original coordinate system such that it is aligned with
this tensor.

\section{Results}\label{sec:results}

\begin{figure*}
  \includegraphics[width=6in]{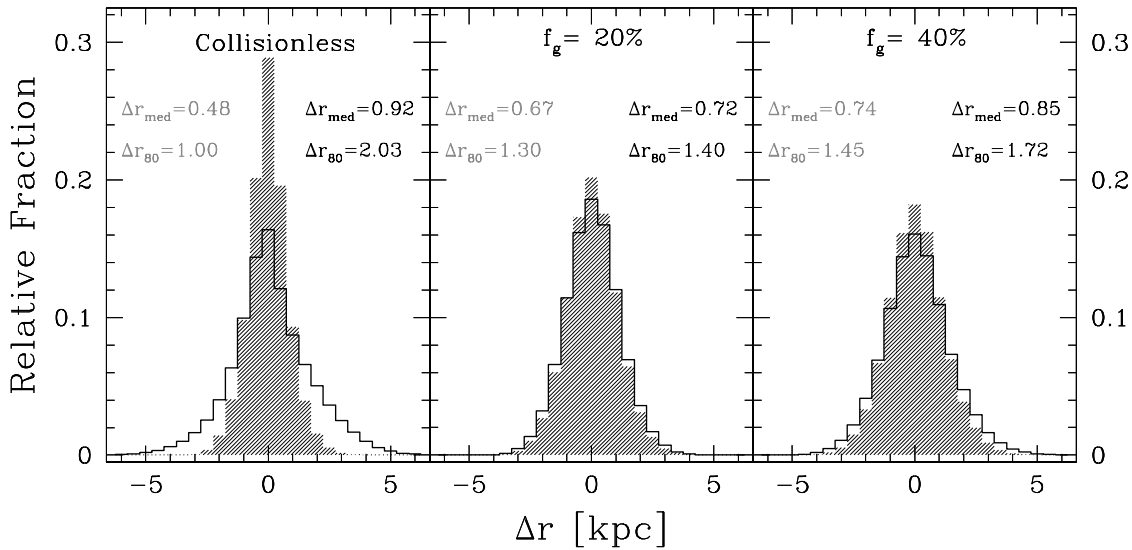}
  \caption{\label{fig:isodisks} The distribution of radial shift
    (\dr$=\rfin-\rinit$) for all disk particles in each of the six
    isolated disk simulations. \mdr\ and \drt\ specify the median and
    $80$th percentile distance traveled, respectively. Gray values
    refer to the hatched histograms; black values are associated with
    the solid black line in each panel. We report the fraction of all
    particles in non-overlapping $500 \pc$ bins of \dr. The two
    collisionless simulations (left panel) illustrate the effect of
    disk structure on radial mixing, \ie, there is more radial
    migration in the smaller scale height (\zd{200}, black line)
    disk than in its thicker counterpart (\zd{400}, hatched
    histogram). The four hydrodynamical simulations are in the middle
    ($f_g=20\%$) and right ($f_g=40\%$) panels. In both of these
    panels, the gray, hatched histograms represent the simulations
    with higher star formation efficiency (\sfe$=0.05$); those with
    \sfe$=0.01$ are shown in black. Both increased gas fractions and
    smaller star formation efficiencies yield greater radial mixing.}
\end{figure*}

Figure~\ref{fig:sims} presents surface density maps of the stellar
distributions of our four collisionless galactic disks with different
initial scale-heights $z_d$. Each simulation exhibits a distinctive
combination of morphological features that could affect radial
migration. The initial smooth disks are unstable to spiral
instabilities arising from the swing amplification of particle shot
noise, an effect that leads to emerging spiral structure as seen in
the final isolated disks \citep[\eg][]{Julian66}. Owing to its smaller
``effective'' Toomre $Q$ stability parameter \citep[\eg,][]{Romeo11},
the \zd{200} disk (hereafter, disks with this initial scale height
will be referred to as ``thin'') develops prominent spiral structure
and a strong bar. Conversely, the \zd{400} disk (hereafter, disks with
\zd{400} will be referred to as ``thick'') has relatively little
spiral structure and does not form a bar in isolation. We emphasize
that the ``thick'', \zd{400} disk is the one in best agreement with
the old, thin disk of the MW, while \zd{200} is too thin (see
Section~\ref{sec:isolated_disks}).

The radial and vertical morphology of the perturbed disks is distinct
from their isolated counterparts. Both perturbed disks develop bars
and are characterized by prominent flaring and much larger scale
heights than those of the isolated disks (K08;
\citealt{Kazantzidis09}). There is evidence that some spiral structure
evident in the isolated disks has been washed out in the simulations
with substructure bombardment: within $10 \kpc$ of the galactic
center, local enhancements of the stellar surface density evident in
the isolated disks are substantially more diffuse after the action of
the infalling satellites. Throughout this section, we will investigate
how the growth and dissipation of spiral structure affect radial
migration.  While we have investigated similar maps in the four
hydrodynamical simulations, we do not show them here as they are
qualitatively similar to their collisionless counterparts. As
expected, however, the magnitude of spiral structure increases as the
gas fraction rises. We note that the differential effect of gas on the
strength of spiral structure we find here is smaller than that
reported in previous numerical investigations of isolated disk
galaxies \citep[\eg,][]{Barnes_Hernquist96}. This is mainly due to
the effect of stellar feedback, which causes the ISM in our
simulations to become turbulent and multi-phase \citep[see
also][]{Stinson_etal06}.

\subsection{Radial Migration}\label{sec:radmix}

We first investigate where particles move throughout each
simulation. Figure ~\ref{fig:isodisks} shows the distribution of
\dr$=\rfin - \rinit$ in the six isolated disk simulations, where
\rfin\ and \rinit\ refer respectively to each particle's final and
initial projected distance from the galactic center.  Each panel
contains the \dr\ distribution, the median $|\dr|$ (denoted \mdr), and
$80^{th}$ percentile $|\dr|$ (denoted \drt) for two of the six
isolated galaxies in our simulation suite.

The isolated, collisionless simulations (left panel) clearly
demonstrate that disk scale height and gas content affect the radial
migration process. The thin disk's \dr\ distribution (black line) is
considerably broader than the thick disk's (gray hatch). Both
\mdr$=0.91\kpc$ and \drt$=2.03\kpc$ of particles in the thin disk are
approximately double those found in the thick disk.

The remaining panels of Figure~\ref{fig:isodisks} compare the \dr\
distributions of the hydrodynamical simulations of our sample. We
examine four isolated disks with two initial gas fractions: $f_g=20\%$
(middle panel) and $f_g=40\%$ (right panel). The two histograms in
each panel represent star formation efficiencies (\sfe) of $1\%$
(solid line) or $5\%$ (gray, hatched histogram).  Gas has a strong
impact on particle radial migration: \mdr\ increases from $0.47\kpc$
in the collisionless case (all hydrodynamical simulations are of
``thick'' disks) to $\sim 0.70\kpc$ when $f_g=20\%$ and $\sim
0.80\kpc$ when $f_g=40\%$. The extent of radial migration is less
dependent on star formation efficiency; there are only minor
differences amongst each pair of histograms in the middle and right
panels. However, in both panels, \mdr\ and \drt\ are highest when
\sfe$=1\%$. In these four hydrodynamical simulations, the
time-averaged gas content of the disk is directly correlated with the
percentage of particles that migrate and their typical displacement
relative to their formation radius.

Decreasing the scale height of the stellar disk, increasing the
fraction of the initial disk mass in gas, and lowering the
star-formation efficiency all increase the midplane density of the
disk. This enhanced density increases the coherence and self-gravity
of spiral structure, supporting it against the dissipative nature of
individual particles' random motions \citep[as described by,
\eg,][]{Toomre77}. The correlation of spiral structure strength with
the fraction of particles that migrate away from their birth radii and
the median migratory distance traversed suggests that the SB02
mechanism, intimately linked with spiral waves, plays a major role in
the migration of particles in the isolated systems.  The symmetric
shapes (to within $0.4\%$ about $0\kpc$) of the \dr\ distributions are
also consistent with migration via the SB02 mechanism. Resonances
between the bar and spiral structure could also influence the
timescale for migration in the thin disk models \citep{Minchev10}, but
the \zd{400} models do not develop bars, regardless of whether they
include gas.

\begin{figure}
  \includegraphics[width=3.4in]{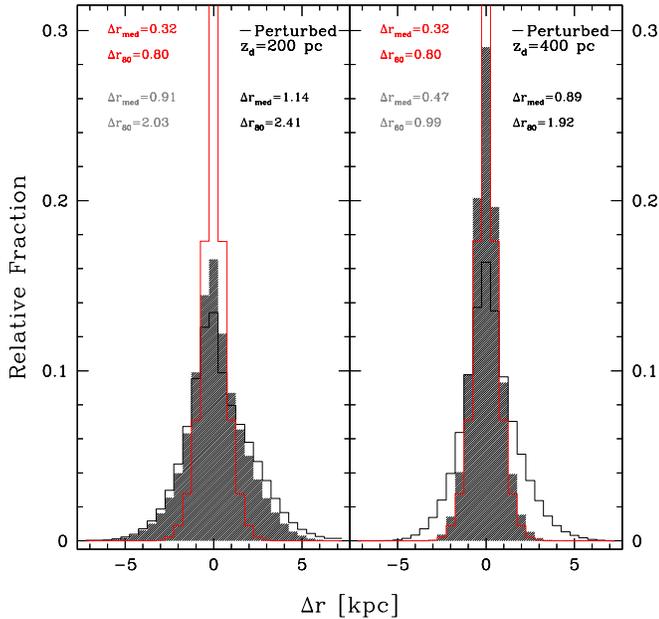}
  \caption{\label{fig:perdisks}The \dr\ distribution for all disk
    particles in the four collisionless simulations. Each histogram
    show the fraction of particles in non-overlapping $500 \pc$ bins
    of \dr. The isolated (hatched, gray histogram) and perturbed
    (black line) \zd{200} (left panel) and \zd{400} (right panel)
    disks are shown. The perturbed disks show greater dispersion in
    \dr\ compared to isolated galaxies with the same geometry. The red
    histograms denote the expected \dr\ distribution from epicyclic
    motion alone, given the initial orbital configurations of the
    particles in each disk. \mdr\ and \drt\ are labeled and
    color-coded for the three distributions in each panel.}
\end{figure}

Figure~\ref{fig:perdisks} compares the \dr\ distributions of the
isolated collisionless disks to those of their perturbed
counterparts. The \mdr\ and \drt\ of each distribution are labeled and
color-coded to match the corresponding histogram. In each panel, the
most sharply peaked histogram (red) shows the expected \dr\
distribution from epicyclic motion alone, which we compute given the
particle's initial phase space coordinates. We model the epicycle as a
simple harmonic oscillator about the particle's guiding center radius
(\rg) with an amplitude set by its initial radial energy
($E_\mathrm{r}$) \citep[Chapter 3, ][]{Binney08}. The energy
associated with the circular component of the particle's orbit
($E_\mathrm{circ}$) is set by \rg, which is the radius of a circular
orbit with angular momentum equal to the midplane component of the
particle's angular momentum. The radial energy of the orbit is
$E_\mathrm{r} = E_\mathrm{tot} - E_\mathrm{z} - E_\mathrm{circ}$ where
$E_\mathrm{z}$ is the vertical energy component and $E_\mathrm{tot}$
is the total energy . As $E_\mathrm{r}$ sets the amplitude of the
epicycle, we can solve for the position of the particle as function of
time with respect to its guiding center radius.  We choose two random
phases of the epicycle oscillation, designating the first as \rinit\
and the second as \rfin, and compute \dr. The resulting ``baseline''
distribution is shown in red and is a result of observing the two
random phases (akin to our ``initial'' and ``final'' snapshots) of the
initially elliptical orbits. It does not take into account the
potential heating of these initial orbits and the subsequent increase
in amplitude of their radial motion (blurring).  The shape of the
distribution does not change if we change our random seed or average a
larger number of phases to obtain \rfin\ and \rinit. For the isolated
thick disk (right panel), the \dr\ distribution from full dynamical
evolution is only marginally broader than the expected baseline
distribution, suggesting that these stars are not heated with time and
continue to follow their initial orbits. However, satellite
bombardment broadens the \dr\ distribution dramatically, nearly
doubling both \mdr\ and \drt, necessitating guiding center
modification of a substantial fraction of the orbits.

For the thin disk, isolated evolution produces a much broader \dr\
distribution than the baseline distribution, demonstrating the impact of the
bar\footnote{Bars can increase the eccentricity of particles,
contributing to the ``blurring'' of the disk, and their potential
resonance overlap with spiral structure can induce migration
\citep[see][]{Minchev10}.} and spiral structure in this more unstable
disk. In this case, satellite bombardment only slightly increases the
width of the \dr\ distribution, despite the strong impact on disk
structure that is visually evident in Figure~\ref{fig:sims}. We
suspect that the small net difference between these two histograms
reflects a cancellation between two competing effects of satellite
bombardment. Accretion events heat the stellar disk and thereby
suppress the development of spiral structure, thus reducing the level
of SB02 migration. However, the accretion events also induce radial
mixing directly via their dynamical perturbations. The \dr\ histograms
of both perturbed disks, while still approximately symmetric, are
noticeably more asymmetric than those of the isolated disks, with
$2\%$ ($3\%$) more particles moving outwards than inwards in the thin
(thick) cases, compared to $\leq 0.4\%$ for the isolated models.

\begin{figure}
  \includegraphics[width=3.4in]{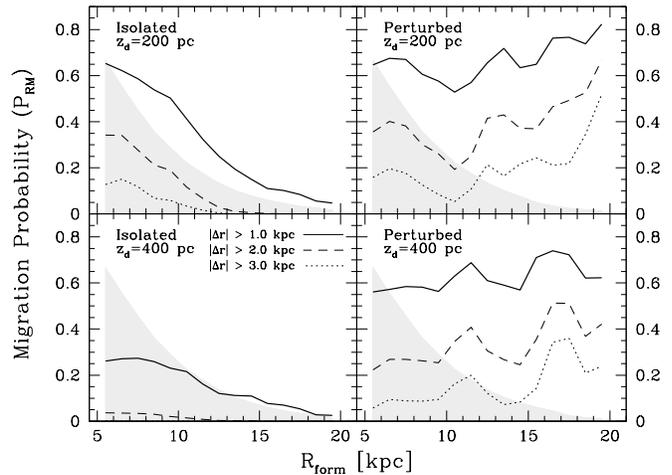}
  \caption{\label{fig:mass_dr} The fraction of particles that migrate
    more than $1.0\kpc$ (solid line), $2.0\kpc$ (long dashed line),
    and $3.0\kpc$ (dotted line) as a function of formation
    radius. Results are binned such that the migration probability is
    calculated for particles in non-overlapping $1.0\kpc$ wide
    annuli. Lines connect the migration fraction in each bin
    (x-coordinates are bin centers, from $5.5$ to $19.5\kpc$). The
    gray regions are the radial mass profiles of the initial disks
    normalized such that the total mass contained in the first annulus
    equals $2/3$ on this scale. The migration probability follows the
    mass distribution in the isolated disks but is anti-correlated
    with mass in the perturbed disks.}
\end{figure}

Figure ~\ref{fig:mass_dr} presents clear evidence that
satellite-induced migration is distinct from that in the isolated
experiments, acting in different environments from either epicyclic
blurring or spiral-induced churning. In each panel, solid, dashed, and
dotted lines show the fraction of particles at each birth radius
\rform\ that migrate by more than $|\dr|= 1.0$, $2.0$, or $3.0\kpc$,
respectively. Shaded regions show the scaled radial surface density
profile of the initial disk. For both isolated disks, the migration
probability decreases outwards and approximately traces the surface
density profile. This behavior is similar to that assumed in the
chemical evolution models of \citet{Schonrich09}, who parametrized the
probability that a star migrates as proportional to the mass
surrounding it. However, for both perturbed disks the probability of
migration is flat or increasing outwards beyond \rform$=10\kpc$, where
the disk potential weakens, and it definitely does not trace the mass
distribution. To better understand this difference between satellite-
and spiral-induced radial mixing, we now investigate how changes in
angular momentum and energy are correlated with change in radius.

%%%%%%%%%%%%%%%%%%%%%%%%%%%%%%%%%%%%%%%%%%%%%%%%%%%%%%%%%%%%%%%
\subsection{Orbital Characteristics} \label{sec:orbits}
%%%%%%%%%%%%%%%%%%%%%%%%%%%%%%%%%%%%%%%%%%%%%%%%%%%%%%%%%%%%%%%

\begin{figure*}
  \includegraphics[width=6.5in]{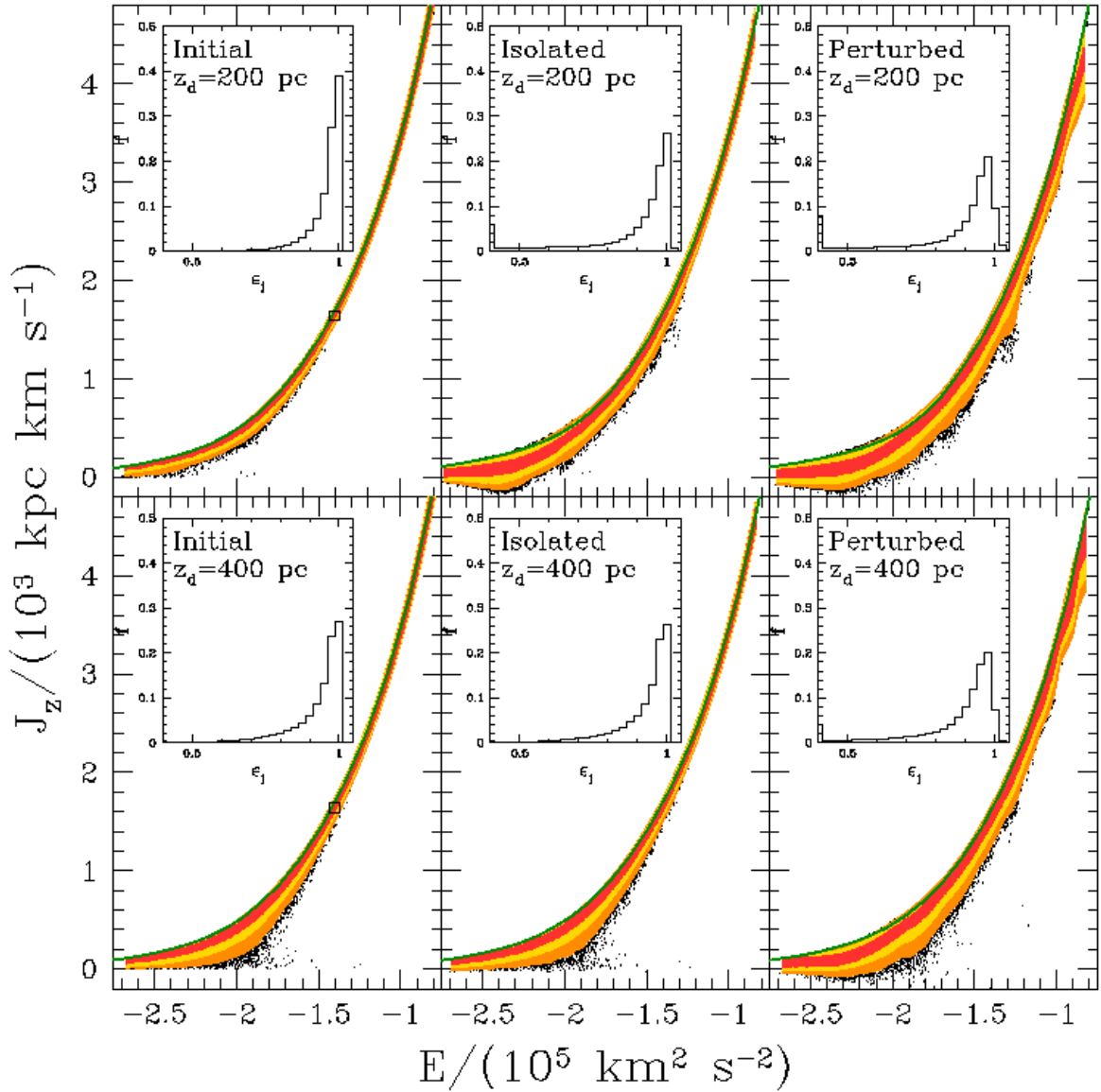}
  \caption{\label{fig:lindblad}Lindblad diagrams for all collisionless
    simulations. Each column shows the initial (left); final, isolated
    (middle); and final, perturbed (right) particle angular momentum
    projected along the $z$ axis ($J_z$, \kpc \kms) vs. specific
    energy ($E$, \kmssq). The red, yellow, and orange regions
    encompass $68\%$, $95\%$, and $99\%$, respectively, of particles
    centered on the median $J_z$ as a function of $E$. $J_z > 99\%$
    outliers are plotted as points. The green line denotes the
    theoretical angular momentum - energy curve for circular
    orbits. Each panel includes a histogram illustrating the
    distribution of circularity (\cir) for all disk particles (see
    text for details). The isolated thin disk shows evolution towards
    slightly non-circular orbits. There is little change in \cir\ in
    the isolated \zd{400}\ case. Simulations with satellite
    bombardment show a pronounced shift towards less circular
    orbits. For reference, the total energy of midplane orbits at $2$,
    $5$, $10$, and $15\kpc$ from the galactic center is approximately
    $-2.0$, $-1.6$, $-1.2$, and $-1.0$ \kmssq, respectively. }
\end{figure*}

Total energy and angular momentum are the classic two-dimensional
integrals of motion \citep{Binney08}. Figure~\ref{fig:lindblad} shows
Lindblad diagrams for the two initial and four final states of the
collisionless simulations, plotting particle specific angular momentum
projected along the axis of symmetry ($J_z$) versus specific binding
energy ($E$).  Particles are grouped into 50 distinct linear bins of
$E$. We calculate the median $J_z$ in each bin. Red, yellow, and
orange regions connect the $68^{th}$, $95^{th}$, and $99^{th}$
percentile $J_z$ intervals (centered on each median $J_z$),
respectively, across all energy bins.  The $1\%$ most discrepant
particles in $J_z$ for a given $E$ are plotted as individual
points. Using each disk's rotation curve, we plot $J_z$ and $E$ for
circular orbits in the midplane of the disk (green line). Here,
$J_z=v_c(r)\times r$ , where $v_c(r)$ is the circular velocity at
radius $r$ and $E=\frac{1}{2}v^2_c(r) + \Phi(r)$, where $\Phi(r)$ is
the potential in the disk midplane at radius $r$. Particles on a
circular orbit have the maximum $J_z$ allowed given their energy.

By construction, particles are initially (left column of
Figure~\ref{fig:lindblad}) on nearly circular orbits with a small radial
velocity component. The red region, falling close to the circular
orbit curve in both initial disks, confirms that $\sigma_\mathrm{v_r}$
is small. Particles significantly displaced from the green curve in
the initial states either have extreme $v_\mathrm{r}$ or are on highly
inclined orbits ($J_z$ is projected along the $z$ axis). Due to this
inclination effect, the initial thick disk has a lower median $J_z$ as
a function of $E$ than the thin disk despite having the same
$v_\mathrm{r}$ distribution.

In the isolated thin disk (upper middle panel of
Figure~\ref{fig:lindblad}), the range of $J_z$ grows relative to the
initial values at every energy $E$. The change is largest at lower
energies. Recall from Figure~\ref{fig:sims} that the isolated thin
disk develops a bar, which is composed of particles on radial orbits
with relatively low $J_z$, in the most bound region of the disk. Thus,
this relatively large change towards less circular orbits at low
energies can be associated with bar formation.  In contrast, the
distribution of $J_z$ in the isolated thick disk is basically
equivalent to its initial state. In the isolated thick disk there is
no bar, relatively little spiral structure develops, and we find the
least radial migration among the four simulations. Ignoring the region
of the Lindblad diagram influenced by members of the bar, the isolated
disk diagrams suggest that the extensive radial migration seen in the
thin disk and associated with guiding center modification decreases
the angular momentum of particles.

The two perturbed disks (right column of Figure~\ref{fig:lindblad})
show larger changes in their Lindblad diagrams. Bar formation in both
simulations can explain the significantly lower median $J_z$ at lower
energies. At higher energies, corresponding to less bound particles
further out in the disk, both disks show a much larger dispersion in
$J_z$ than their isolated counterparts.  Additionally, there is
substructure in the Lindblad diagrams (groups of relatively low $J_z$
in a narrow range of energy) not seen in the isolated disks. Satellite
bombardment in the perturbed disk simulations has a qualitatively
discernible impact on the angular momentum distribution of the disk.

To quantify changes in $J_z$, we introduce the circularity (\cir)
quantity \citep[\eg, ][]{Abadi03}. For a particle of energy $E_i$ and
angular momentum $J_i$ in the $z$-direction, we define circularity as
$\epsilon= J_i/J_\mathrm{circ}(E_i)$, the ratio of $J_i$ to the
specific angular momentum the particle would have if it were on a
circular orbit with energy $E_i$ (obtained using the circular orbit
curve). Circular orbits have $\epsilon=1$ and radial orbits have
$\epsilon=0$. Negative circularities correspond to retrograde orbits.
We note that this method is formally different from that of
\citet{Abadi03}. The circular orbit curve may not represent the
highest $J_z$ for a given $E$ due to shot noise in the potential or
gravitational potential asymmetries. Strictly defining the maximum
$J_z$ in each energy bin (as in \citet{Abadi03}) results in the same
qualitative trends discussed later. Our method benefits from the use
of a mathematically constructed and reproducible rotation curve and
systematically decreases \dcirc$=\epsilon_\mathrm{f} -
\epsilon_\mathrm{i}$ at the $0.01$ level. The inset of each panel in
Figure~\ref{fig:lindblad} shows the circularity distribution of each
simulation.

\begin{figure}
  \includegraphics[width=3.4in]{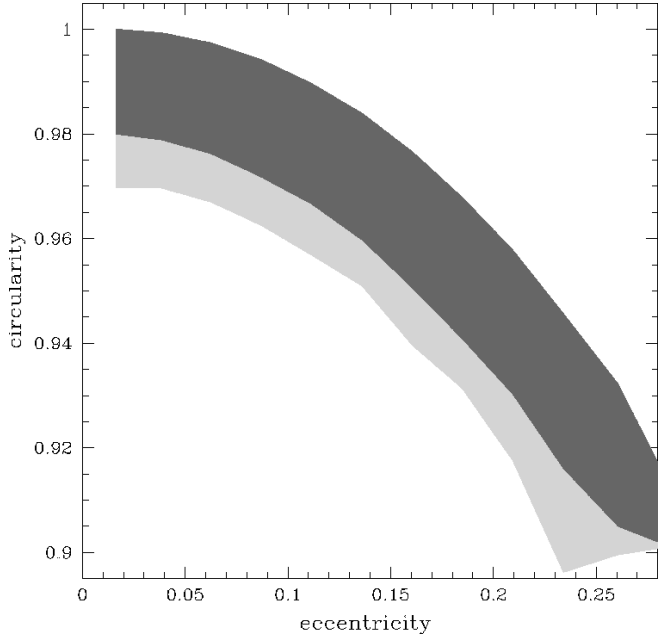}
  \caption{\label{fig:ecc_circ} Possible values of circularity as a
  function of eccentricity in the solar annulus ($7\geq r \geq 9\kpc$)
  of the \zd{400} disk. The top edge of the dark region represents
  orbits in the midplane of the disk. The dark region encompasses the
  $95^{th}$ percentile of particle circularity in the solar annulus of
  each simulation; the bottom edge of the light region corresponds to
  the $99^{th}$ percentile circularity. }
\end{figure}

Figure~\ref{fig:ecc_circ} plots the relation between circularity and
eccentricity for orbits near the solar radius ($8.0\kpc$) in the
\zd{400} disk. Circularity is related to eccentricity via the
particle's radial and vertical energies as well as its orbital
inclination (since $J_z$ is a projected quantity). Orbits in the
midplane of the disk have the highest circularity for a given
eccentricity. Shaded regions in Figure~\ref{fig:ecc_circ} show the
$95^{th}$ and $99^{th}$ percentile circularity as a function of
eccentricity at the solar radius. For reference in interpreting
Figure~\ref{fig:lindblad} and subsequent figures, it is worth noting
that changes of $\sim 0.02$ in \cir\ typically correspond to quite
noticeable changes in eccentricity. The boundaries of these regions
are not smooth due to binning and small number statistics for
initially high eccentricity particles.

Returning to Figure~\ref{fig:lindblad}, we see that the two isolated
disks show markedly different evolution of their circularity
distributions. In the isolated thin disk, the fraction of stars with
\cir\ $\sim1$ (rightmost bin) drops from $0.38$ to $0.26$, and the
median \cir\ drops from $0.980$ to $0.955$. Particles in the bar
predominantly populate the newly formed low circularity tail. In the thick isolated disk, on the other hand, the
\cir\ distribution is nearly identical to the initial disk's,
with median \cir\ dropping only $0.002$. This lack of evolution
in the circularity distribution is fully consistent with CR
scattering. As described by SB02, individual stars may exchange
angular momentum across CRs (churning) while the overall distribution
would remain unchanged. However, our results from Section~\ref{sec:radmix} show
that the Delta R distribution of this model, which reflects individual
particles and their orbits, is nearly identical to that expected from
observing the particles' initially elliptical orbits. The circularity
distribution of the isolated, thick disk combined with its Delta R
distribution imply that, on average, individual particle guiding
centers are not significantly modified.

The circularity distributions of the perturbed disks are demonstrably
altered from their isolated counterparts but are similar to one
another. The most common circularities are now in the range $0.96\leq
\cir \leq 0.98$ in both disks, with a decrease for \cir$>0.98$. The
thin disk starts with more \cir\ $\sim 1$ particles because of its
smaller orbital inclinations, and its circularity distribution evolves
more strongly, with median \cir\ dropping from $0.980$ to $0.936$
vs. $0.963$ to $0.929$ for the perturbed thick disk. Notably, the
perturbed thin and thick disks evolve to similar \dr\
(Figure~\ref{fig:perdisks}) and \cir\ distributions despite starting
with different scale heights.

\begin{figure}
  \includegraphics[width=3.4in]{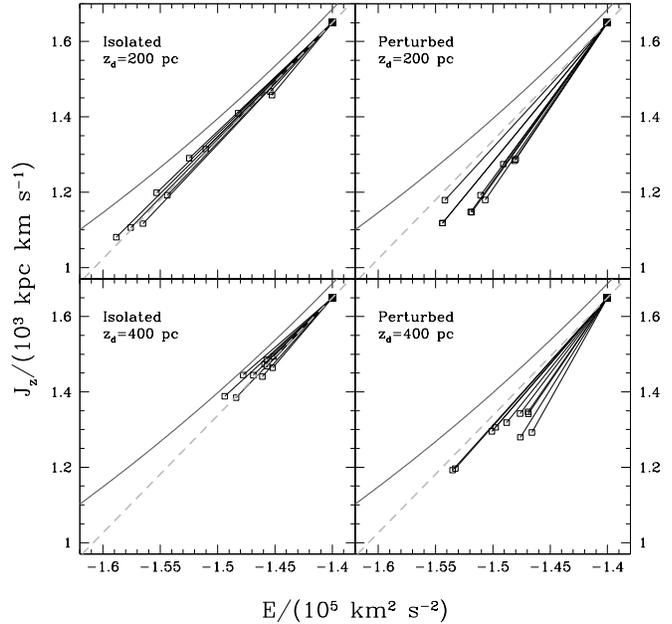}
  \caption{\label{fig:lindzoom} The initial and final $E$, $J_z$ pairs
for ten randomly selected particles in each simulation with initial
energy $-1.405\leq E_\mathrm{i} \leq -1.395 \times 10^5$ \kmssq\ and
angular momentum $1.645\leq J_{z, \mathrm{i}} \leq 1.655 \times 10^3$
\kpc \kms\ (large, filled square). For clarity we require that plotted
particles lose at least $0.05 \times 10^5$ \kmssq\ in energy during
the simulation.  The final $E$,$J_z$ pairs of the ten particles are
plotted as open squares. Lines connect the initial and final $E, J_z$
points of each particle. The $E, J_z$ curve populated by circular
orbits in the initial state of each simulation is indicated by the
thick black curve. The dashed line is tangent to the circular orbit
curve at the initial energy of all ten particles.}
\end{figure}

Figure~\ref{fig:lindzoom} tracks the changes of selected individual
particles in the ($E$, $J_z$) space of the Lindblad diagram. While $E$
and $J_z$ are not individually conserved in the presence of a
non-axisymmetric perturbation, the Jacobi invariant $I=E-\Omega_b J_z$
is, where $\Omega_b$ is the pattern speed of the perturbation, assumed
to be static and small (SB02; \citealt{Sellwood10}). If $\Delta I= 0$,
then $\Delta J_z / \Delta E \approx \Omega_b$. The SB02 mechanism
operates at the corotation resonance of the star/particle and the
spiral wave, requiring that $\Omega_b=\Omega_\mathrm{rot}$. Thus, in
the galaxy, particles should move parallel to the line that is tangent
to the circular orbit curve at their binding energy prior to
scattering. In other words, the SB02 mechanism requires that changes
in energy be accompanied by changes in angular momentum that preserve
the orbital shape, modulo differences in the slope of the circular
orbit curve over the range $[E_\mathrm{i}, E_\mathrm{f}]$ of a given
particle.

Figure~\ref{fig:lindzoom} zooms in on the area of the Lindblad diagram
designated by the two small boxes drawn on the initial state diagrams
in Figure~\ref{fig:lindblad}. We randomly select ten particles within
the same initial $E$ and $J_z$ range (filled, dark squares in
Figure~\ref{fig:lindzoom}) and plot their final $E$ and $J_z$. We
require $E_\mathrm{f}- E_\mathrm{i} = \Delta E < -0.05 \times 10^5$
\kmssq\ to ensure that the individual tracks are visible. In the
isolated thin disk (top left), most particles move parallel to the
circular orbit curve tangent (dashed line); this relationship between
$\Delta J_z$ and $\Delta E$ is consistent with corotational resonant scattering (equation 2, SB02). When particles lose energy in the isolated thick disk
(bottom left), their $J_z$ typically remains closer to the circular
orbit curve than in the isolated thin disk. The changes in energy and
angular momentum are relatively small, indicating that the guiding centers
are modified to a modest extent \citep{Binney08}. Note that these particles are in the top $1\%$ of $|\Delta J_z|$; most particles in this particular experiment have $\Delta J_z \approx 0$.  The randomly selected particles in
both perturbed disks have slopes $\Delta J_z/ \Delta E$ that are
steeper than the slope of the tangent to the circular orbit curve at
corotation. This distinct coupling of $\Delta E$ and $\Delta J_z$,
combined with our results concerning the migration probability as a
function of \rform\ (Section ~\ref{sec:radmix}), offer compelling
evidence that migration in the perturbed disks is driven, at least in
part, by a mechanism that does not operate in the isolated
systems. Figure~\ref{fig:lindzoom} shows that the orbital
characteristics of many particles in the perturbed disks are modified
in a fashion inconsistent with either epicyclic motion or a single
spiral wave scattering event.

\begin{figure}
  \includegraphics[width=3.4in]{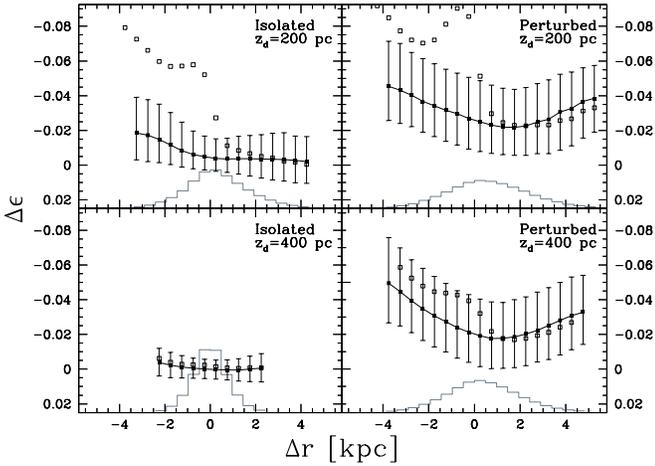}
  \caption{\label{fig:dr_dcirc} Median change in circularity \dcirc\
    as a function of \dr\ for the entire disk (open squares) and for
    particles with $\rfin, \rinit > 4.0\kpc$ (filled
    squares). Particles are sorted into non-overlapping $0.5\kpc$ bins
    of \dr. Error bars mark the $25^{th}$ and $75^{th}$ percentile
    \dcirc\ in each bin. Histograms at the bottom of each panel
    represent the relative fraction of particles in each bin for the
    case $\rfin, \rinit > 4.0\kpc$. }
\end{figure}

We now examine the correlation between migration and changes in
orbital properties. Since the metallicity of star-forming gas
increases with time and decreases with radius, any such correlations
also imply observable correlations between the present orbital
parameters and metallicities of stars as a function of age and
Galacto-centric radius. Figure~\ref{fig:dr_dcirc} shows the median
change in circularity, $\dcirc= \epsilon_\mathrm{f} -
\epsilon_\mathrm{i}$, as a function of radial migration distance \dr\
in the four collisionless simulations (open squares). In the isolated
thick disk; changes in circularity are tiny (median $|\dcirc| \leq
0.005)$ at any \dr; Figure~\ref{fig:lindzoom} shows that particles in
this simulation stay close to the circular velocity curve even if they
change energy. In the other three simulations, particles with negative
\dr\ show substantial drops in circularity (typical median $\dcirc <
-0.06$), which are almost certainly associated with bar formation. The
bars that form in these three simulations increase the orbital
eccentricities of their members. To focus on behavior in the disk
proper, the filled squares in each panel show the median \dcirc\
vs. \dr\ for those particles that start and end the simulation at $r
\geq 4 \kpc$, beyond the extent of the bar. Error bars mark the
inter-quartile range ($25^\mathrm{th}$ to $75^\mathrm{th}$ percentile)
at each \dr.

In the isolated thin disk, particles that migrate inwards (\dr$<0$)
experience a modest decrease in circularity, stronger for more
negative \dr\ (and $\Delta E$), consistent with the tracks shown in
Figure~\ref{fig:lindzoom}. Particles that migrate outwards (\dr$>0$)
increase their energy and typically traverse areas of the Lindblad
diagram with relatively little change in the slope of the circular
orbit curve. Following the tangent to the circular orbit curve,
particles will not significantly change their circularity in such a
scenario (note the relative lack of low circularity particles at $E >
-1.5 \kmssq$ in Figure~\ref{fig:lindblad}).  The perturbed disks show
a decrease in median circularity at every \dr, and a much wider
inter-quartile range indicating a greater range of orbital
inclinations and eccentricities. Circularity drops more strongly for
particles that have experienced strong radial migration, either inward
or outward. The minimum in these curves is slightly offset to positive
\dr\ because dynamical heating slightly puffs up the disk radially,
decreasing the potential at a given radius, thereby increasing its
total energy, moving particles to the right on the Lindblad diagram ,
and lowering the circularity for particles with \dr$=
0$. Figure~\ref{fig:dr_dcirc} indicates that stars with anomalous
chemistry for their age and current position should have
preferentially more eccentric orbits. The trend is smaller than the
inter-quartile range, but similar in magnitude.

\begin{figure}
  \includegraphics[width=3.4in]{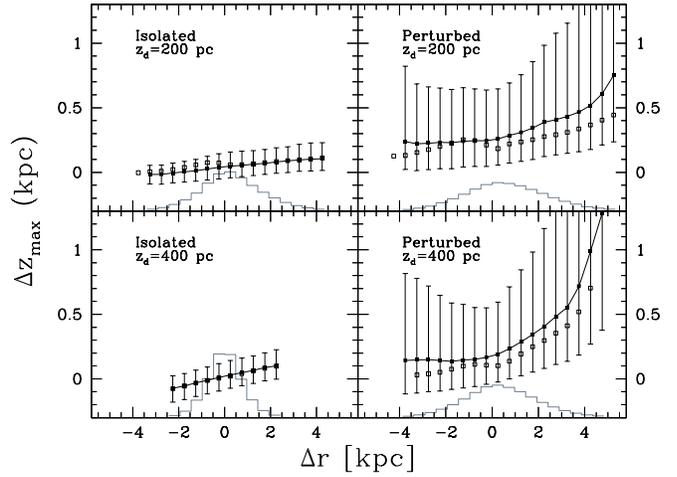}
  \caption{\label{fig:dr_dzmax} Like Figure~\ref{fig:dr_dcirc}, but
  the change in maximum vertical displacement, \dzm, is plotted in
  place of the change in circularity. Open squares show all disk
  particles, filled squares show those with \rfin, \rinit\ $> 4 \kpc$,
  and error bars mark $25^{th}$ and $75^{th}$ percentile at a given
  \dr.}
\end{figure}

\citet{Schonrich09} and \citet{Loebman10} discuss the possible role of
radial migration in producing thick disks like the ones observed in
the Milky Way \citep{Gilmore89} and other edge-on galaxies
\citep{Dalcanton02}. Figure~\ref{fig:dr_dzmax} is similar to
Figure~\ref{fig:dr_dcirc}, but instead of $\Delta\epsilon$ it plots
the change in vertical energy, quantified by the maximum distance \zm\
that a particle can reach from the disk plane. We compute \zm\
approximately from each particle's vertical velocity component
assuming that the final potential of each simulation is static, namely
\zm$=|z| + \frac{v_z^2}{4\pi G\Sigma(r)}$ where $z$ is the vertical
position of the particle at the final output, $v_z$ is the velocity
along the $z$ axis, $G$ is the gravitational constant, and $\Sigma(r)$
is the surface density of the disk at radius $r$ assuming all the mass
of the disk is in the midplane. The change in \zm\ is
\dzm$=z_{\mathrm{max,f}} - z_{\mathrm{max,i}}$ where the subscripts
$i$ and $f$ refer to the initial and final simulation snapshots,
respectively.

The two isolated simulations show a shallow linear trend between \dzm\
and \dr\ (open squares for all particles; filled squares for those
with \rinit,\rfin$ > 4 \kpc$). When particles move outwards (inwards)
through the disk, they experience a weaker (stronger) gravitational
potential, thus increasing (decreasing) \zm. However, changes in \zm\
are small, less than $300$ pc even when we consider the quartile range
at the extremes of \dr. The perturbed disks, by contrast, show larger
changes in the median \zm\ and a dramatic increase in the
inter-quartile range at all \dr. K08 show that the perturbed $400$ pc
disk develops a two-component vertical structure in quantitative
agreement with the observed thin/thick disk structure of the MW. The
perturbed $200$ pc disk increases its scale height (to $\approx 500
\pc$ at the solar annulus), but it can still be described by a single
component model. This vertical heating by satellite perturbations is
evident in Figure~\ref{fig:dr_dzmax}. Particles that have large
positive \dr\ have the largest increase in \zm, which is plausibly a
consequence of moving outwards to regions of lower disk surface
density and thus weaker vertical restoring force. For \dr\ $<0$, the
trend of median \dzm\ with \dr\ is approximately flat, suggestion a
cancellation between the effects of increased $\Sigma(r)$ at smaller
$r$ and direct satellite-induced heating of those particles with the
largest excursions. Figure~\ref{fig:dr_dzmax} implies that stars with
high metallicity for their age and present location should have
preferentially larger \zm, though the scatter is larger than the
trend.

\subsection{Solar Annulus} \label{sec:soln}

\begin{figure}
  \includegraphics[width=3.4in]{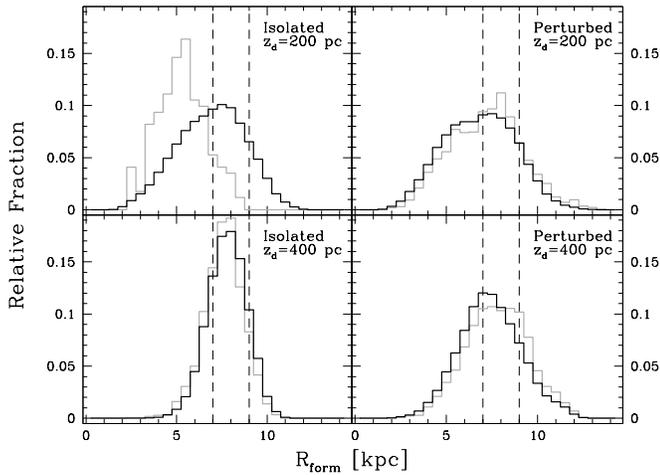}
  \caption{\label{fig:rform} The distribution of formation radius
    (\rform, in \kpc) for all particles that are in the solar annulus
    ($7\kpc \leq r \leq\ 9\kpc$) at the end of each
    simulation. Histograms report the fraction of solar annulus
    particles emigrating from non-overlapping $500 \pc$ annuli in
    \rform. Particles within the dashed lines remained in the solar
    annulus throughout the simulation. Both thin disks and the
    perturbed thick disk show a broad range of \rform\ at the solar
    annulus. The gray histogram in each panel is the \rform\
    distribution for those particles that are in the solar annulus and
    at large heights above the plane ($1.0 < |z| < 1.5\kpc$) in the
    final simulation output.}
\end{figure}

The solar neighborhood is easier to study than other regions of the
Galaxy, since high-precision spectroscopy is easier for brighter stars
and parallax and proper motion measurements are more accurate at
smaller distances. Some of the most detailed chemo-dynamic surveys,
such as the GCS \citep{Nordstrom04} and the
Radial Velocity Experiment \citep[RAVE, ][]{Steinmetz06} concentrate
on the solar neighborhood. In this section, we repeat some of our
earlier analysis specifically for stars that reside in the solar
annulus ($7\kpc \leq \rfin \leq 9\kpc$) at the end of each
simulation. This focus on the solar annulus also removes much of the
impact of the bars that dominate evolution of the inner disk ($r < 3
\kpc$) in three of our simulations, though some particles from the bar
region can migrate as far as the solar radius, and resonances between
the bar and spiral structure may increase migration frequency
\citep{Minchev10}.

Figure~\ref{fig:rform} shows the radius of formation (\rform)
distribution of particles residing in the solar annulus, marked by the
vertical dashed lines. Only the isolated thick disk simulation
predicts a final solar annulus dominated by stars born in the solar
annulus, with tails extending $1$--$2 \kpc$ on either side. The broad
\dr\ distributions of the other three simulations show that their
stars migrate to the solar annulus from a wide range of formation
radii. The global \dr\ (Figure~\ref{fig:perdisks}) and solar annulus
\rform\ distributions of the isolated and perturbed thin disks are
remarkably similar, despite the differences in migration mechanisms
discussed in Section~\ref{sec:orbits}. In the isolated thin disk,
$32\%$ of solar annulus particles originated at \rform$\leq 6 \kpc$
and $4\%$ at \rform$\geq 10 \kpc$. Corresponding numbers for the
perturbed thin disk are $36\%$ and $3\%$. The \rform\ distribution of
the perturbed thick disk is slightly narrower, but it still broad with
respect to the isolated thick disk.

\begin{figure}
  \includegraphics[width=3.4in]{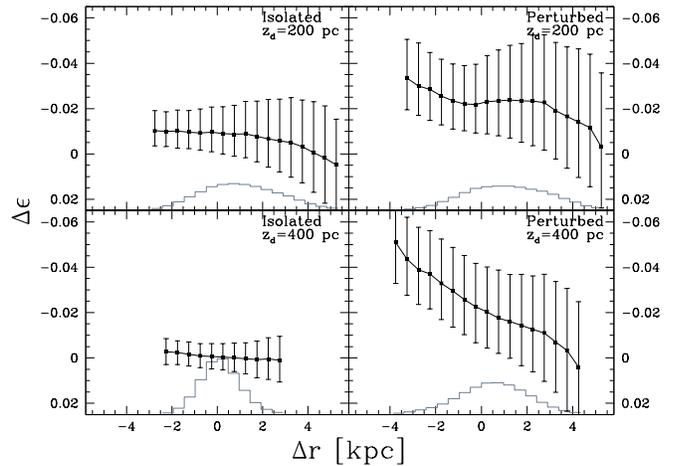}
  \caption{\label{fig:dcirc_dr_soln} The change in circularity \dcirc\
    as a function of \dr\ for particles ending the four collisionless
    simulations in the solar annulus ($7\kpc \leq \rfin\ \leq\
    9\kpc$).  Particles are sorted into non-overlapping $0.5\kpc$ bins
    of \dr. We plot the median (squares) and the $25^{th}$ and
    $75^{th}$ percentile (error bars) \dcirc\ in each bin. Histograms
    at the bottom of each panel represent the relative fraction of
    particles in each bin.}
\end{figure}

Figure~\ref{fig:dcirc_dr_soln} shows the correlations between \dcirc\
and \dr\ for particles that end in the solar annulus. Consistent with
results for the full disk (Figure~\ref{fig:dr_dcirc}), the solar
annulus particles in the isolated thick disk show no significant
change in median \cir\ regardless of \dr. In the isolated thin disk,
the range of \dcirc\ is much larger, with a modest decrease in median
\cir. The median \dcirc\ drops at large positive \dr\ because
particles move to less inclined orbits as they migrate outwards and
disk heating has slightly modified the galaxy's circular velocity
curve, allowing outward moving particles to potentially increase
circularity. In the perturbed thick disk, there is a strong and nearly
linear trend between \dcirc\ and \dr. Particles that migrated to the
solar annulus from the outer disk have experienced substantial drops
in circularity, while particles migrating from the inner disk show
only modest decreases. The range of \dcirc\ is large at all
\dr. Results for the perturbed thin disk are intermediate between
those of the isolated thin and perturbed thick disks: quasi-linear
trends at large $|\dr|$ but a flat plateau at intermediate
$|\dr|$. This behavior is plausibly a consequence of two different
mechanisms contributing to migration, with spiral-induced mixing
dominating at intermediate \dr\ and satellite-induced mixing
dominating at the extremes.

\begin{figure}
  \includegraphics[width=3.4in]{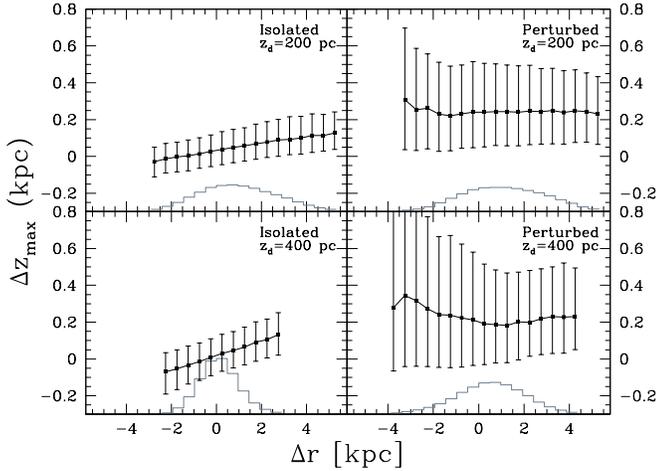}
  \caption{\label{fig:dzmax_dr_soln} Like
  Figure~\ref{fig:dcirc_dr_soln}, but the change in vertical
  displacement \dzm\ is plotted in place of the change in
  circularity.}
\end{figure}

Figure~\ref{fig:dzmax_dr_soln} shows the solar annulus correlations of
\dzm\ with \dr, analogous to Figure~\ref{fig:dr_dcirc} for the full
disk. For the two isolated disks there is a clear linear trend of
median \dzm\ with \dr\ as expected from the arguments in
Section~\ref{sec:orbits}: particles that migrate outward move to a
region of lower disk surface density, so if their vertical velocities
are not systematically changed by the radial migration they will
attain higher \zm. Both perturbed disks show signs of the strong
vertical heating induced by satellite accretion. The median \dzm\ is
$\geq 0.2 \kpc$ almost independent of \dr, and the range of \dzm\ is
much larger than in the isolated disks. Since inwardly migrating
particles experience a higher vertical potential at \rfin\ than
\rinit, they must experience more vertical heating than outwardly
migrating particles to keep the \dzm\ trend flat.

Returning to Figure~\ref{fig:rform}, gray histograms represent the
\rform\ distributions of particles that end the simulations in the
solar annulus at high $z$, $1.0 < |z| < 1.5\kpc$. In the isolated
thick disk and both perturbed disks, the \rform\ distribution of high
$z$ particles resembles that of all solar annulus particles. In these
three models, therefore, selecting high- $z$ particles does not
isolate a population with atypical radial migration. In the isolated
thin disk, on the other hand, the \rform\ distribution of high-$z$
particles is strongly skewed towards low formation radii, with a peak
at \rform$= 5 \kpc$. Here, only initially hot particles that migrate
significantly outwards via churning or interactions with the bar and
experience a weaker potential have enough vertical energy to overcome
the local restoring force and reach high $z$. Only $0.1\%$ of the
solar annulus is at high $z$ in the isolated thin disk (truly the tail
of the initial vertical velocity dispersion); this number rises to
$3.2\%$ and $7.5\%$ in the thin and thick perturbed disks,
respectively. The radial migration mechanisms in the perturbed disks
ensure that even the high $z$ population of the solar annulus comes
from a broad range of \rform. Measurements of the age-metallicity
relation for high-$z$ stars could be a valuable diagnostic for
distinguishing models of radial migration and vertical heating.

\section{Summary and Discussion}\label{sec:conclusion}

Observations suggest that radial migration plays an important role in
the chemical evolution of the MW disk \citep[][]{Wielen96,
Schonrich09}. SB02 described a mechanism by which stars at
corotational resonance with spiral waves can scatter, producing large
changes in guiding center radius while keeping stars on nearly
circular orbits. Previous simulations have shown that migration over
several kiloparsecs can occur in isolated disks grown by smooth
accretion \citep{Roskar08a} and that encounters with satellites can
also induce migration \citep{Quillen09}. Here we have carried out a
systematic investigation of a variety of simulations to characterize
the role of stellar disk properties, gas fractions, and satellite
perturbations in producing radial migration. Most importantly, our
suite of simulations includes experiments with a level of satellite
bombardment expected in $\Lambda$CDM models of galaxy formation, which
earlier investigations \citep[K08, ][]{Kazantzidis09} have shown to
produce vertical and in-plane structure resembling that seen in the
MW.

For disks evolved in isolation, the degree of migration correlates
with the degree of disk self-gravity, susceptibility to bar formation,
and spiral structure. The collisionless stellar disk with \zd{400},
chosen to match that of the thin disk in the MW, has no bar and
minimal spiral structure. It exhibits limited radial migration; the
median value of radial change $|\rfin\ - \rinit|$ is \mdr$ = 0.47
\kpc$, consistent with the level expected from epicyclic motion
(Figure~\ref{fig:perdisks}). The collisionless \zd{200} disk is much
more unstable, develops a bar and spiral structure
(Figure~\ref{fig:sims}), and has a higher degree of radial migration,
with \mdr$=0.91 \kpc$ and \drt$=2.03 \kpc$. The presence of gas is a
catalyst for radial migration in the \zd{400} case. Both \mdr\ and
\drt\ increase in models with larger gas fractions or lower star
formation efficiency (which consumes gas more slowly). The galactic
bar strongly influences, and perhaps dominates, the resultant
individual particle dynamics in the inner galaxy. Substantial radial
migration, consistent with the SB02 mechanism or being
satellite-induced, occurs in the outer portion of the disk in all of
our collisionless experiments except for the isolated \zd{400}
disk. In the \zd{200} disk, stars that finish the simulation in the
solar annulus ($7 < R < 9 \kpc$) have \mdr$=1.36 \kpc$, \drt$=2.68
\kpc$.

Adding satellite perturbations dramatically increases radial migration
in the \zd{400} disks, raising \mdr\ from $0.47 \kpc$ to $0.89 \kpc$
globally and from $0.60 \kpc$ to $1.15 \kpc$ in the solar
annulus. Perturbations do not change the \dr\ distribution so
drastically in the \zd{200} disks, but there are other indications
that the nature of radial migration is different. In the isolated
galaxies, migrating particles in the outer disk follow tracks in $E,
J_z$ space that parallel the tangent to the local circular velocity
curve, as expected for the SB02 mechanism (see
Section~\ref{sec:orbits}). However, inwardly migrating particles in
the perturbed disks lose more angular momentum for a given change in
energy, causing migration associated with heating and thus different
from the SB02 mechanism in this respect. The relationship between
$\Delta E$ and $\Delta J_z$ found in the perturbed experiments is
broadly more consistent with scattering at LR (see SB02),
but a detailed decomposition of the modes in the disk and satellites
(beyond the scope of this paper) is necessary to confirm LR scattering
on a particle by particle basis. The radial distribution of migrating
particles presents an even clearer distinction, one with important
implications for chemical evolution models
(Figure~\ref{fig:mass_dr}). In the isolated disks, the probability
that a particle with a formation radius \rform\ undergoes significant
migration ($|\dr|> 1 \kpc$) is approximately proportional to the disk
surface density at \rform; \ie, migration follows mass. In the perturbed disks, the probability of migration
is flat or increasing with radius, so a much larger fraction of
migration comes from the tenuous outer disk. Particles in the outer
disk are overall more likely to migrate relative to the inner disk as 1) they
feel a weaker potential and are thus more susceptible to direct
heating by satellites and 2) their circular frequencies make them more
likely to resonantly interact with satellites and migrate. Thus,
satellite-induced migration patterns are distinct from those produced
by purely secular evolution.

Satellite bombardment heats the disk both vertically and
radially. Compared to the isolated disks, the perturbed disks
experience a greater drop in median circularity (hence growth of
median eccentricity) during their evolution, and individual particles
experience a wider range of circularity changes. The circularity
change is moderately correlated with radial migration distance and
direction, though the trend is smaller than the inter-quartile range
at a given \dr\ (Figure~\ref{fig:dr_dcirc}). If we restrict our
analysis to the solar annulus, the correlation between \dr\ and
\dcirc\ is more significant. In particular, particles in the solar
annulus of the \zd{400} disk that migrate $2$--$4 \kpc$ inwards
experience substantial drops in circularity, while those that migrate
the same distance outwards nearly maintain their initial circularity
(Figure~\ref{fig:dcirc_dr_soln}). The large inward excursions from the
outer disk, driven by satellite perturbations, systematically remove
angular momentum from particle orbits.

As shown by K08, vertical heating by satellite bombardment produces
(in the case of the \zd{400} disk) a two-component vertical structure
in quantitative agreement with the thin and thick disk profiles
observed in the MW. In the isolated disks, we find the expected trend
that as particles migrate outwards, they experience a weaker vertical
potential and increase their vertical energy (characterized by \zm,
the maximum distance from the plane that a particle can reach given
its current location and velocity). However, in our simulations, this
effect is not sufficient to produce a second, ``thick-disk'' component
in the isolated systems even if they have substantial radial
migration. The satellite perturbations increase the median and range
of \zm\ considerably at all \dr. The resulting overall trend of \dzm\
with \dr\ is fairly flat (especially at a fixed \rfin, such as the
solar annulus, see Figures~\ref{fig:dr_dzmax}
and~\ref{fig:dzmax_dr_soln}), suggesting that systems with low
vertical velocity dispersion and subjected to weaker potentials (as in
the outer disk) are more susceptible to vertical heating (similar to
the radial heating trends seen in Section~\ref{sec:radmix}). Particles
with \rform\ interior \emph{or} exterior to a given \rfin\ can be
found at significant distances above the plane in the perturbed
disks, while increases in \zm\ are smaller in the isolated systems and
rely on particles moving outwards (Figure~\ref{fig:rform}).

Our results confirm earlier findings that spiral structure development
in isolated disks \citep{Roskar08a, Loebman10} and perturbations by
satellites \citep{Quillen09} can produce significant radial mixing of
stellar populations while retaining reasonable orbital structure for
disk stars. They strongly support the view (\citealp{Wielen96};
SB02; \citealp{Schonrich09}) that radial mixing is an essential ingredient
in understanding the chemical evolution of the MW and disk galaxies in
general. While a combination of metallicity and age can be used to
estimate a star's formation radius given a chemical evolution model,
the uncertainties in this approach (including the difficulty of
estimating ages for typical stars in a spectroscopic survey) will make
it difficult to reconstruct formation radius distributions for
observed stars, even in the solar neighborhood. However, our analysis
provides theoretical guidance for chemical evolution models that
incorporate radial mixing \citep[\eg][]{Schonrich09} and suggestions
for the correlations one might search for between chemical abundances
and orbital properties (though the predicted trends are fairly
weak). We regard our perturbed \zd{400} disk as the most relevant
simulation, as it includes the satellite accretion events expected in
\LCDM\ and produces a final vertical structure similar to that
measured in the MW. In this simulation $41\%$ of particles in the $7
\kpc < R < 9 \kpc$ solar annulus were ``born'' in that annulus, $20\%$
migrated there from $R < 6 \kpc$, and $7\%$ migrated there from $R >
10 \kpc$.

Radial mixing and orbital dynamics changes are sensitive to several
different aspects of disk modeling, as shown by the variety of our
results. Robust predictions should therefore come from calculations
that self-consistently include star formation, chemical enrichment,
gas accretion, and accretion events, and the interplay among these
elements. We will move in this direction with our future
simulations. Giant spectroscopic surveys such as SEGUE, RAVE, APOGEE,
LAMOST, and HERMES offer an extraordinary opportunity to unravel the
formation history of the MW, and they offer an exciting new challenge
to theoretical models of galaxy formation.

\section*{Acknowledgments}
We thank the anonymous referee for a careful and constructive
report. We acknowledge useful discussions with Mario Abadi, Julianne
Dalcanton, Rok Ro{\v s}kar, Ralph Sch\"onrich, Scott Gaudi, Chris
Kochanek, Victor Debattista, Simone Callegari, Tom Quinn, and James
Wadsley. S.K. is funded by the Center for Cosmology and Astro-Particle
Physics (CCAPP) at The Ohio State University.  This work was supported
by NSF grants AST $0707985$ and AST $1009505$, and by an allocation of
computing time from the Ohio Supercomputer Center
(http://www.osc.edu). This research made use of the NASA Astrophysics
Data System. D.W. acknowledges the hospitality of the Institute for
Advanced Study and the support of an AMIAS membership during part of
this work.

%\bibliography{rm_mnras}

\label{lastpage}

\end{document}